\documentclass[12pt]{iopart}
\usepackage{iopams}
\expandafter\let\csname equation*\endcsname\relax
\expandafter\let\csname endequation*\endcsname\relax
\usepackage{amssymb}
\usepackage{amsbsy}
\usepackage{color}
\usepackage{graphicx}
\usepackage{color}
\usepackage{amsmath}

\usepackage{multirow}

\usepackage{ulem}

\begin{document}

\title{Extreme-mass-ratio burst detection with TianQin}

\author{Hui-Min Fan$^1$, Shiyan Zhong$^{2}$, Zheng-Cheng Liang$^1$ ,  Zheng Wu$^1$, Jian-dong Zhang$^1$,Yi-Ming Hu$^{1\dagger}$}

\address{$^1$ TianQin Research Center for Gravitational Physics, 
Sun Yat-Sen University, Zhuhai 519082, China}

\address{$^2$ Yunnan Observatories, Chinese Academy of Sciences, 396 Yang-Fang-Wang, Guandu District, 650216, Kunming, Yunnan, China}

\ead{zhongsy@ynao.ac.cn} 
\ead{huyiming@mail.sysu.edu.cn} 

\begin{indented}
\item[] September 73, 2022
\end{indented}

\begin{abstract}
The capture of compact objects by massive black holes in galaxies or dwarf galaxies will generate short gravitational wave signals, called extreme-mass-ratio bursts (EMRBs), before evolving into extreme-mass-ratio inspirals. Their detection will provide an investigation of the black hole properties and shed light on astronomy and astrophysics. In this work, we investigate the detection number of the TianQin observatory on EMRBs. Our result shows that TianQin can detect tens of EMRBs events during its mission lifetime. For those detected events, we use the Fisher information matrix to quantify these uncertainties in the inference of their parameters. We consider the possible network of TianQin+LISA, and study how a network can improve parameter estimation. The result shows that, for most sources, the CO mass, the MBH mass, and the MBH spin can be determined with an accuracy of the order $10^{-1}$ and the sky localization can be determined with an accuracy of 10 square degree.  We further explore the gravitational wave background generated by those unsolved EMRBs and conclude that it is about $10^6$ times weaker than TianQin's sensitivity and thus it can be ignored.
\end{abstract}

\vspace{2pc} \noindent{\it Keywords}: {Galaxy nuclei (609) --- Supermassive black holes (1663) --- Stellar dynamics (1596) --- N-body simulations (1083) --- Tidal disruption (1696)---gravitational waves---
EMRB---TianQin }

\section{Introduction}\label{sec:intro}

The space-based gravitational-wave (GW) detectors TianQin, planned to be launched in the 2030s, are aiming to detect  GW sources in the mHz band\cite{TianQin:2015yph,TianQin:2020hid}. Within this frequency band, there are plenty of sources including Galactic ultra-compact binaries\cite{Hu:2018yqb,Huang:2020rjf,Brown:2020uvh,Kremer:2017xrg,Korol:2017qcx}, coalescing massive black holes (MBHs)\cite{Wang:2019ryf,Feng:2019wgq,Ruan:2021fxq,Shuman:2021ruh,Katz:2019qlu}, the low-frequency inspirals of stellar-mass black holes\cite{Liu:2020eko,Klein:2022rbf,Buscicchio:2021dph,Ewing:2020brd,Toubiana:2020cqv}, the extreme-mass-ratio insprials\cite{Fan:2020zhy,Zhang:2022xuq,Wardell:2021fyy,Lynch:2021ogr,Isoyama:2021jjd,Vazquez-Aceves:2022dgi},and the stochastic GW backgrounds \cite{Liang:2021bde,Renzini:2022alw,LISACosmologyWorkingGroup:2022kbp,Boileau:2020rpg}. Detecting those sources has great significance, as they will help to study the formation and evolution of  compact objects (COs) and MBHs\cite{DeLillo:2022blw,Wang:2019kaf,Amaro-Seoane:2022rxf}, allow for testing gravitational theories in strong field region and for checking the validity of the black hole no-hair theorem\cite{Zi:2021pdp,Shi:2019hqa,Barsanti:2022ana,Rahman:2021eay}. Apart from these traditional sources, TianQin can also be used to search for GW bursts and unforeseen sources.  One of those GW burst signals are generated by COs captured by MBHs, called EMRBs \cite{Han:2020dql}\cite{Berry:2013ara}\cite{Berry:2013poa}\cite{Berry:2012im}, which are formed during the early evolution stage of EMRIs. Here, we do an exploration of the detectability of TianQin on EMRBs.  

Galactic nuclei generally host a massive black hole at its center\cite{Kormendy:1995er,Magorrian:1997hw,Gebhardt:2002js,Ferrarese:2004qr} and feature a dense structure of stars and compact objects\cite{Alexander:2005jz,Schodel:2014wma}. Relaxation processes in such a high-density environment occasionally force stars and compact objects onto extremely eccentric, low angular momentum orbits, resulting in a close encounter with the central MBH\cite{Yunes:2007zp}\cite{Hopman:2006fc}\cite{Rubbo:2006dv}.  Each time the CO gets close to the pericenter of the orbit, an EMRB signal is generated .  Due to the emission of GWs, the orbit of the CO circularizes and decays gradually until forming a continuous source called extreme-mass-ratio inspirals (EMRIs)\cite{AmaroSeoane:2007aw}. The detection of EMRBs can provide useful information to the study of astronomy and astrophysics. It can help us understand how the MBHs are formed, gain a glimpse into the past of the MBHs' host galaxies, and study the COs formation and evolution history\cite{genzel2010galactic}.

To estimate the detection number of TianQin on EMRBs, we need the corresponding astrophysical model to get the EMRB population properties. In this paper, we use the recently published mass function given by E. Gallo and A. Sesana\cite{Gallo:2019fcz} to explore the MBH distribution and estimate the rate of EMRBs by performing N-body simulations. EMRBs has large mass ratios and strong field background, therefore perturbation theory can be used to calculate their waveform\cite{Hughes:2021exa,Babak:2006uv,Katz:2021yft,Berry:2012im}. Assuming the CO as a point mass, its trajectory can be characterized by the geodesic motion in the MBH background. We use the quadrupole approximation here since the accuracy for the EMRB is sufficient while it greatly reduce the calculation costs. This method also known as numerical kludge (NK) and have a detailed description in \cite{Babak:2006uv}.



We choose a SNR of 10 as the detection threshold for EMRBs in TianQin, we expect those EMRB signals which have signal-to-noise ratio (SNR) above this value can be detected by TianQin. Our result shows that TianQin can detect tens of EMRB events during its mission lifetime. For those EMRB sources detected,  our result shows that although most of the EMRB source parameters can't be determined very well, in the best case scenario, the CO mass, the MBH mass, and the MBH spin can be determined with an accuracy of the order $10^{-2}$ and the sky location can be determined with an accuracy of 10 square degree.  We further consider a combined detection by TianQin and LISA. The result shows that the CO mass, the MBH mass, and the MBH spin of most sources can be determined with an accuracy of the order $10^{-1}$ and the sky location can be determined with an accuracy of 10 square degree. For the most precisely determined source, the CO mass, the MBH mass, and the MBH spin can be determined with an accuracy of the order $10^{-3}$ and the sky location can be determined with an accuracy of $10^{-2.5}$ square degree. Last, We explore the GWB generated by EMRBs, finding that it is about $10^6$ times weaker than TianQin's sensitivity and thus it can be ignored.

This paper is organized as follows, In Sec. II, we have a description of the distribution of EMRB populations. In Sec. III, we illustrate the EMRB waveform calculation method and the response function of TianQin to EMRBs. In Sec. IV, we show the detection result. In Sec. V, we present our conclusions.

\section{EMRBs distribution}\label{paralist}

The dynamical processes leading to EMRB formation and the MBH population are two key points to getting the EMRB distribution. In section \ref{rateEMRB}, we estimate the rate of EMRBs, which take place in the vicinity of MBHs residing in either galaxies or dwarf galaxies, by performing direct $N$-body simulations. We describe the details of the $N$-body simulations in section \ref{Nsim} and study the dependence of the EMRB rate on the MBH mass in section \ref{ratePerM}.  In section \ref{Mdist}, we described the MBH mass function, which is used to illustrate the MBH population. Then, we extract Montecarlo realizations of the EMRB events and get their distribution.
\subsection{The rate of EMRBs}\label{rateEMRB}

The waveforms of EMRB depend on many parameters (see equation \ref{paras} in the next section), among which the $N$-body simulations could provide the mass of the MBH ($M$), the mass of the CO ($m$) and pericentric distance ($r_{\rm p}$) at which an EMRB takes place. Therefore, the event rate obtained from the simulation is formally written as $\Gamma_{\rm EMRB}(M,m,r_{\rm p})$, which can be further split into 
\begin{equation}
\Gamma_{\rm EMRB}(M,m,r_{\rm p})=\dot{N}_{\rm EMRB}(M) f(m,r_{\rm p}). 
\end{equation}
The quantity $\dot{N}_{\rm EMRB}(M)$ is the overall rate of EMRB events that comprises all $(m,r_{\rm p})$ combinations, while $f(m,r_{\rm p})\mathrm{d}m \mathrm{d}r_{\rm p}$ gives the fraction of events happened in the parameter space element around the point $(m,r_{\rm p})$ and  satisfies the normalization condition $\int f(m,r_{\rm p})\mathrm{d}m \mathrm{d}r_{\rm p} = 1$.

The simulated rate of EMRBs varies significantly from model to model, therefore, what we adopted in the expression of $\Gamma_{\rm EMRB}(M,m,r_{\rm p})$ is $\langle\dot{N}_{\rm EMRB}(M)\rangle$, the mean value of $\dot{N}_{\rm EMRB}(M)$ averaged over all the simulated models. Meanwhile, the rate of MBH-CO coalescence, $\dot{N}_{\rm coal}(M)$, is relatively consistent across all the models, and its dependence on $M$ could be derived from the loss cone theory. Hence we use the mean value of $\dot{N}_{\rm coal}(M)$ as the calibration point and express $\langle\dot{N}_{\rm EMRB}(M)\rangle = \kappa\langle\dot{N}_{\rm coal}(M)\rangle$.

\subsubsection{$N$-body simulation}\label{Nsim}

The initial positions and velocities of the particles (representing the stars and COs) in $N$-body model star clusters are sampled from the Plummer distribution following the method of \cite{aarseth1974comparison}. Due to the large fluctuation of $\dot{N}_{\rm EMRB}(M)$ in individual models, a large ensemble of models is simulated to cover the cases as many as possible, so that a statistically meaningful value of $\langle\dot{N}_{\rm EMRB}(M)\rangle$ can be obtained. We have chosen 10 different random seeds to initialize the particle distribution. Initially, the mass of the cluster is $M_{\rm c}=[M]$ and all the particles have the same mass, i.e. they are the star particles having the mass of $(1/N) [M]$, where $N$ is the particle number and $[M]$ is the mass unit in the $N$-body model. Next, 0.5 percent of the particles are substituted for CO particles, whose masses are sampled from a power law distribution $n(m)\propto m^{-2}$ in the mass range of $(1/N) [M]$--$(50/N) [M]$ (i.e. the physical mass of COs are taken from the range of $1 M_{\odot}$--$50 M_{\odot}$). We use another 10 random seeds to sample the masses of the COs, hence in total, we have 100 model clusters.

A central MBH is presented as an external potential placed at the center of the model cluster. The mass of the MBH is chosen as $5\%$ of the cluster mass. A coalescence radius $r_{\rm coal}$ is assigned for the central MBH, so that a CO coalesces with the MBH if the distance $d$ between them falls below $r_{\rm coal}$. For Schwarzschild BH and near parabolic orbit, the coalescence radius $r_{\rm coal} = 4 r_{\rm g}$.  For Kerr BH and near parabolic orbit, the coalescence radius $r_{\rm coal} =  r_{\rm g}$\cite{Stein:2019buj,Glampedakis:2002ya}. To be more conservative, we choose $4 r_{\rm g}$ as the coalescence radius. The EMRB emission is significant when $r_{\rm p}\leq 10 r_{\rm g}$ \cite{Berry:2013poa}. Hence in the simulation, an EMRB event occurs when the pericentric distance of the CO to the MBH falls into the range of $r_{\rm coal} \leq r_{\rm p} \leq 2.5 r_{\rm coal}$. The particle number in the $N$-body model (see below) is orders of magnitude lower than the number of stars in the realistic cluster, hence a boosted $r_{\rm coal}$ is adopted in the $N$-body model to ensure a sizable number of coalescence events. We set $r_{\rm coal} = 4\times10^{-5} [L]$, where $[L]$ is the length unit of the $N$-body model and is corresponding to the virial radius of the star cluster.

Direct $N$-body simulation is computationally expensive, since the number of arithmetical operations scales with $N^2$, though the scaling relation can be improved to $\propto N^{1.613}$ with some special arrangements for the timesteps\cite{berczik2011high}. We choose a particle number $N=32$K (1K=1024) for all the 100 models for the following considerations. First, with this $N$ the number of COs in each model is 163 and could ensure that at least one CO has a mass close to $50 M_{\odot}$. Second, given the large ensemble of models to be simulated and the limited computational resources, this choice of $N$ could bring the total simulation time down to an acceptable level.

The simulations are performed with the \texttt{Nbody6++GPU} code\cite{wang2015nbody6++}\cite{huang2016performance} with a new subroutine named \texttt{accretion} which deals with the accretion of stars and COs onto the MBH\cite{Panamarev:2018bwq}. The MBH do not gain mass from the accreted CO or star, in order to prevent the artificial fast growth of the MBH mass due to the limited mass resolution (see \cite{zhong2014efficient} for example). We have modified this subroutine so that it can record the EMRB events as well.

All the models are simulated for $1000 [T]$, where $[T]$ is the time unit in the $N$-body model. This time corresponds to roughly 3 relaxation timescales of the model cluster. During the simulation, we observe $26.95 \pm 6.66$ MBH-CO coalescence events averaged over the 100 models, while the average number of EMRB events is 3129. We did not give the value of standard deviation because the distribution of $N_{\rm EMRB}$ is very asymmetric about the mean value and has a long tail extending to almost $15000$. The resultant ratio of $\langle\dot{N}_{\rm EMRB}\rangle$ to $\langle\dot{N}_{\rm coal}\rangle$ is $\kappa = 116.1$.

The fractional distribution function $f(m,r_{\rm p})$ is shown in Fig.\ref{fig:fmrp}

\begin{figure}
\centering
\includegraphics[width=\columnwidth,clip=true,angle=0,scale=1.1]{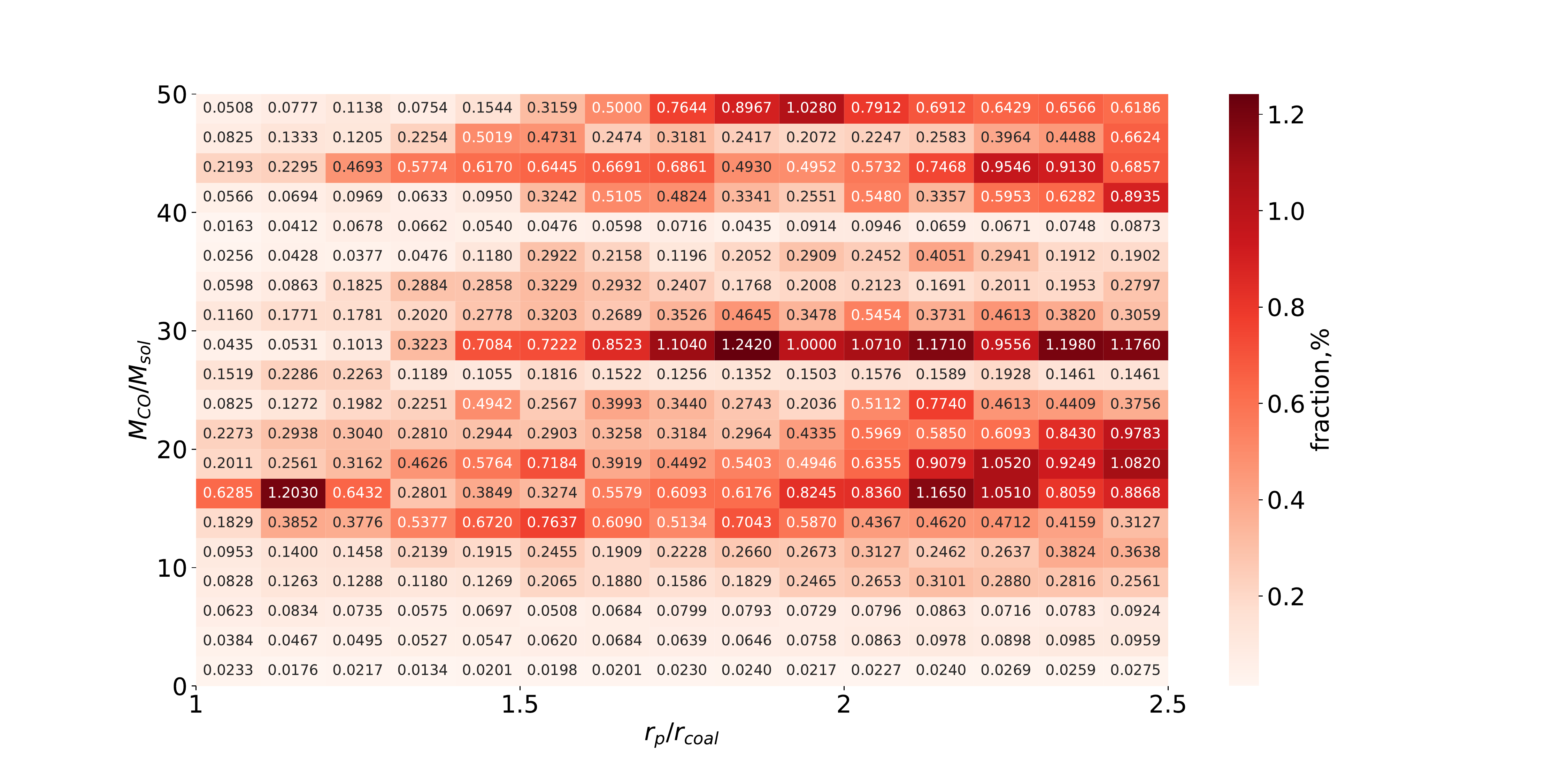}
\caption{The fractional distribution function $f(m,r_{\rm p})$, where $M_{\rm CO}$ corresponding to the CO mass $m$ and $r_{\rm p}$ corresponding to the pericenter distance of the CO orbit. }
\label{fig:fmrp}
\end{figure}

\subsubsection{The dependence of EMRB rate on the mass of MBH, $\langle\dot{N}_{\rm EMRB}(M)\rangle$}\label{ratePerM}

The MBH-CO coalescence events are similar to the tidal disruption events (TDEs) in the sense that the intruding object is ``accreted" by the MBH after the events, and the COs and stars are delivered to the MBH via similar dynamical processes. In the spherical symmetric star cluster adopted in this work, this is done via the two-body scattering process occurring near the apocenter of the orbit and predominantly perturbing the orbital angular momentum. Hence, we could compute the event rate of MBH-CO coalescence based on the loss cone theory \cite{frank1978effects}, which shows that the event rate depends on the number of objects ($N$) residing in the star cluster and the corresponding ``accretion" radius ($r_{\rm acc}$) (also see \cite{Panamarev:2018bwq,kennedy2016star,zhong2014super,Baumgardt:2004zu}).

We briefly review the scaling relation for the rate of TDEs, then apply this relation to scale the rate of MBH-CO coalescences from $N$-body models to realistic systems. The mass growth rate of the MBH caused by TDEs is estimated as\cite{kennedy2016star}\cite{Baumgardt:2004zu}
\begin{equation}
\dot{M} = \frac{\rho(r_{\rm crit}) r_{\rm crit}^3}{t_{\rm rlx}(r_{\rm crit})}
\label{Mdot}
\end{equation}
\noindent
where $\rho(r)$ is the stellar density, $t_{\rm rlx}(r)$ is the relaxation time scale, and $r_{\rm crit}$ is the critical radius where the loss cone repopulating process is balanced with the consumption of stars inside the loss cone\cite{frank1978effects}.

The MBH is embedded in a stellar cusp with density profile $\rho(r) = \rho_0 (r_{\rm inf}/r)^s$. Applying the condition that the enclosed stellar mass within $r_{\rm inf}$ equals  $M$, the quantity $\rho_0$ is expressed as a function of $M$ and $r_{\rm inf}$,

\begin{equation}
\rho_0 = \frac{(3-s)M}{4\pi r_{\rm inf}^3}.
\label{rho_0}
\end{equation}
\noindent
Inside the stellar cusp ($r < r_{\rm inf}$), we assume the gravitational potential is dominated by the MBH, hence the velocity dispersion of the stars follows $\sigma^2(r) = GM/r$. Inserting the expressions of the relaxation time scale

\begin{equation}
t_{\rm rlx}(r) = \frac{\zeta \sigma^3(r)}{G^2 \overline{m}_{\rm star} \rho(r) \ln(\Lambda N)},
\label{t_rlx}
\end{equation}
\noindent
and the density profile $\rho(r)$ into equation~\ref{Mdot}, and substituting $\overline{m}_{\rm star}$ with $M_{\rm c}/N$ results in a new expression for $\dot{M}$,

\begin{equation}
\dot{M} = F(s) \frac{M_{\rm c}}{t_{\rm dyn}(r_{\rm inf})}
\frac{\ln(\Lambda N)}{N}
\left( \frac{r_{\rm crit}}{r_{\rm inf}}  \right)^{\frac{9-4s}{2}}.
\label{Mdot-1}
\end{equation}
\noindent
where $t_{\rm dyn}(r_{\rm inf})$ is dynamical timescale at the influence radius, $\zeta$ and $F(s)$ are dimensionless constants irrelevant for the scaling procedure.

Following \cite{kennedy2016star} we derived the expression for $r_{\rm crit}/r_{\rm inf}$,
\begin{equation}
\frac{r_{\rm crit}}{r_{\rm inf}} = \left[ D(s)
\left( \frac{M}{M_{\rm c}} \right)
\left( \frac{r_{\rm acc}}{r_{\rm inf}} \right)
\frac{N}{\ln(\Lambda N)}
\right]^{\frac{1}{4-s}}.
\label{r_crit-r_inf}
\end{equation}
\noindent
where $D(s)$ is a dimensionless constant irrelevant for the scaling procedure. Substituting $\dot{M}$ with $\dot{N}_{\rm acc}(M_{\rm c}/N)$ and $r_{\rm crit}/r_{\rm inf}$ with equation~\ref{r_crit-r_inf} in equation~\ref{Mdot-1} we get

\begin{equation}
\dot{N}_{\rm acc} = F(s) \frac{\ln(\Lambda N)}{t_{\rm dyn}(r_{\rm inf})}
\left[ D(s)
\left( \frac{M}{M_{\rm c}} \right)
\left( \frac{r_{\rm acc}}{r_{\rm inf}} \right)
\frac{N}{\ln(\Lambda N)}
\right]^{\frac{9-4s}{8-2s}}.
\label{Ndot}
\end{equation}

This equation tells the $N$ and $r_{\rm acc}/r_{\rm inf}$ dependence of $\dot{N}_{\rm acc}$,

\begin{equation}
\dot{N}_{\rm acc} \propto
\left( N \frac{r_{\rm acc}}{r_{\rm inf}} \right)^{\frac{9-4s}{8-2s}}
[\ln(\Lambda N)]^{\frac{2s-1}{8-2s}}.
\label{Ndot_scaling}
\end{equation}
Note we have assumed that the $N$-body model and the real system have the same $M/M_{\rm c}$, so this term does not enter the scaling formula.

To simulate the real system with a MBH of mass $10^6 M_6 M_{\odot}$, the particle number should be $N_{\rm star}^{\rm real} = 2.0\times 10^7 M_6$. Such a large number of particles is currently unreachable for direct $N$-body simulations. However, with the help of the scaling relation (equation~\ref{Ndot_scaling}), we could estimate what is the $\dot{N}_{\rm acc}$ in a $N$-body model with $N_{\rm star}^{\rm real}$ particles and realistic ``accretion" radius.

Assuming the MBH resides in a Bahcall-Wolf cusp ($s=7/4$)\cite{bahcall1976star} and takes $\Lambda = 0.4$ \cite{Spitzer1987}, the MBH-CO coalescence rate is computed with the following formula

\begin{equation}
\langle \dot{N}_{\rm coal}^{\rm real}\rangle =
\left( \frac{N_{\rm star}^{\rm real}}{N_{\rm star}^{\rm sim}} \right)^{4/9}
\left( \frac{r_{\rm coal}^{\rm real}/r_{\rm inf}^{\rm real}}
{r_{\rm coal}^{\rm sim}/r_{\rm inf}^{\rm sim}} \right)^{4/9}
\left[ \frac{\ln(0.4 N_{\rm star}^{\rm real})}{\ln(0.4 N_{\rm star}^{\rm sim})}
\right]^{5/9}\times \langle \dot{N}_{\rm coal}^{\rm sim}\rangle.
\label{N_coal}
\end{equation}
\noindent
Applying the $M-\sigma$ relation\cite{schulze2011effect} we find the influence radius in the real system to be $r_{\rm inf}^{\rm real} = 1.09 M_6^{0.54}$ pc\cite{zhong2014super}. While the coalescence radius $r_{\rm coal}^{\rm real} = 1.92\times 10^{-7} M_6$ pc. Thus the ratio $r_{\rm coal}/r_{\rm inf}$ in the real system is $1.76\times 10^{-7} M_6^{0.46}$. 
The influence radius of the model cluster is $0.176[L]$ and we set $r_{\rm coal}^{\rm sim}=4\times 10^{-5} [L]$, so the ratio $r_{\rm coal}/r_{\rm inf}$ in the simulation is $2.27\times 10^{-4}$.

From the simulation we find $\langle \dot{N}_{\rm coal}^{\rm sim}\rangle = 2.695\times 10^{-2} [T]^{-1}$, and the unit of $\langle\dot{N}_{\rm coal}^{\rm real}\rangle$ obtained from equation~\ref{N_coal} is also $[T]^{-1}$. We need the physical value of $[T]$ to express the event rate with the physical time unit (e.g. yr$^{-1}$). $[T]$ is estimated by $\sqrt{[L]^3/(G[M])}$,
where $[M]=2\times 10^7 M_6 M_{\odot}$. The influence radius in the $N$-body model is $r_{\rm inf} = 0.176 [L]$, with the physical value of the influence radius $r_{\rm inf} = 1.09 M_6^{0.54}$ pc, we find $[L] = 6.2 M_6^{0.54}$ pc. As a result $[T] = 5.14\times 10^4 M_6^{0.31}$ yr.

In the above equations, all the quantities are only depending on $M$,
hence we rename $\langle \dot{N}_{\rm coal}^{\rm real}\rangle$ as $\langle \dot{N}_{\rm coal}(M)\rangle$. And the scaling relation for the EMRB rate is

\begin{equation}
\langle \dot{N}_{\rm EMRB}(M)\rangle = \kappa\langle \dot{N}_{\rm coal}(M)\rangle.
\label{Equa:EMRBrate}
\end{equation}


\subsection{MBH mass function}\label{Mdist}
The mass function describing the MBH distribution has many uncertainties due to our current limited knowledge of the MBH evolution. From previous work\cite{Fan:2020zhy}, we know that those EMRBs generated by an MBH with a lower mass are more likely to enter the TianQin detection range. 
As observed MBHs usually have a higher mass value than $10^6M_\odot$\cite{greene2020intermediate},  their mass function is not very suitable to get the MBH mass we are interested in.   Recently, E. Gallo+2019 \cite{Gallo:2019fcz} made an exploration of the local black hole mass lower than $10^6M_\odot$ and gave a mass function describing the MBH within $10^{4}\sim10^{9}M_\odot$, which can be regarded as the most direct description of the mass function that yields the most sensitivity sources for the upcoming spacecraft detectors. In this paper, we applied their work to get the MBH distribution, which can be described as
\begin{equation}
\log\Phi(M)=c_1+c_2\log(M/M_\odot)+c_3\ln(\log(M/M_\odot)),
\label{Equa:Mfunc}
\end{equation}
Where  $\Phi$ is the number of black holes per comoving volume, $M$ is the MBH mass, $c_1=-2.13$, $c_2=-0.098$, and $c_3=-0.00011$.  As EMRB signals are rather weak, we don't expect MBH with luminosity distance exceeding 1Gpc to be detectable by TianQin. Thus, we ignore the mass function evolving with the redshift in this paper.  

The MBHs that have been observed usually have near maximal spin. However, in \cite{Jones:2020nnx}, the authors thought that the high spin values observed are possibly due to observation bias, as high quality, high SNR spectral sources are easier to be detected. By measuring the spin of $\sim$1900 AGN, the authors get the average spin value of 0.62. We adopted their work, assuming the MBH spin  satisfies the Poisson distribution with an average value equal to 0.62.

Based on the result above, we can compute the intrinsic EMRB rate as 
\begin{equation}
\mathcal{R}(M,a,m,r_{\rm p})=\Phi(M)\times{p(a)}\times\Gamma_{\rm EMRB}(M,m,r_{\rm p}),
\end{equation}
where $p(a)$ describes the spin distribution for MBHs. We construct the population of  EMRBs events using Monte Carlo sampling and get 12896 EMRB events during the mission lifetime of TianQin.

\section{EMRB Waveform}\label{sec:wave}

EMRBs are characterized by large mass ratios and high pericenter velocities. An approximate kludge method known as NK\cite{Babak:2006uv} can be used to calculate their waveform\cite{Yunes:2007zp}\cite{Berry:2012im}.  In this method, the CO is considered as a point-like object moving along the geodesics of the MBH and the waveform is constructed using a quadrupole formula. The CO geodesics are parametrized by three physical quantities, which are energy $E$, specific angular momentum along the symmetry axis $L_z$, and Carter constant $Q$. As EMRB signals are short, these three quantities can be regarded as conserved and  we ignore their small change induced by GW emission when passing the pericenter. Especially, EMRBs are radiated from a quite eccentric orbit with $e\sim1$,  this makes the CO have a parabolic-like orbit with energy $E=1$, which greatly simplifies the geodesic equations to get the CO evolution. 

The quadrupole approximation waveform can be written as
\begin{equation}\label{WaveEqua}
h_{ij}=(2/D)(P_{ik}P_{jl}-\frac{1}{2}P_{ij}P_{kl})\ddot{I}^{kl},
\end{equation}
Where $P_{ij}\equiv\eta_{ij}-\hat{n}_i\hat{n}_j$ is the projection operator, $D$ is the source luminosity distance, $I^{ij}(t)=\mu{r^i}(t)r^j(t)$ is the inertia tensor,  $\vec{r}$ is the displacement vector of the CO from the MBH and $\mu$ is the CO mass.

In equation \ref{WaveEqua}, the orbital evolution $\vec{r}$ is described by the Kerr geodesic equations\cite{Berry:2012im}
\begin{equation}\label{geoEqus}
\begin{split}
(r^2+a^2\cos^2\theta)\frac{d\psi}{d\tau}&=\Big(2r_{\rm p}-(r_3+r_4)(1+\cos\psi)+\frac{r_3r_4}{2r_{\rm p}}(1+\cos\psi)^2\Big)^{1/2},\\
(r^2+a^2\cos^2\theta)\frac{d\chi}{d\tau}&=\sqrt{Q+L_z^2},\\
(r^2+a^2\cos^2\theta)\frac{d\phi}{d\tau}&=\Big(\frac{L_z}{Q\cos^2\chi/(Q+L_z^2)}-a+\frac{a}{\Delta}[(r^2+a^2)-L_za]\Big),\\
(r^2+a^2\cos^2\theta)\frac{dt}{d\tau}&={\Big(\frac{(r^2+a^2)^2}{\Delta}-a^2-2raL_z/\Delta+\frac{Q}{Q+L_z^2}\cos^2\chi{a^2}\Big)},
\end{split}
\end{equation}
Where $a$ is the dimensionless spin of the MBH, $\Delta=r^2-2Mr+a^2$, $M$ is the MBH mass, the variables $(\psi,\chi,\phi)$ are substitute for $(r, \theta, \phi)$ using the relationship $r=2r_{\rm p}/(1+\cos\psi)$ and $\cos^2\theta=Q/(Q+L_z^2)\cos^2\chi$. This is beneficial when performing the numerical integration. Beside the pericenter distance $r_{\rm p}$,  ($r_3, r_4$) are the other two roots of  the radial potential $V_r(L_z, Q)$, which have detailed expression in \cite{Babak:2006uv}. 

With the geodesic equations, the $I^{ij}$ are calculated in the $\hat{S}$-based coordinate before performing the projection operation,  which should be transformed to the $\hat{n}$-based coordinate before interacting with the projection operation. The quantity $\hat{S}$ is the  spin  direction $(\theta_K, \phi_K)$ of the MBH relative to the line of sight, while $\hat{n}$ is the traveling direction $(\theta_S, \phi_S)$ of the waves emitted during the EMRB.

We also have ($L_z, Q$) with relationships  $Q=\tan^2\iota\cdot{L^2_z}$ and $V_r(L_z, Q)|_{(r=r_p)}=0$, which allow us to replace  ($L_z, Q$) by ($r_{\rm p}, \iota$), where $\iota$ is the inclination angular. The formula $V_r(r_{\rm p}, \iota)=0$  further determines the values of $r_3$ and $r_4$.  Thus the EMRB waveform is described by the parameters
 \begin{equation}\label{paras}
(r_p, M, m, a, \iota, \theta_S, \phi_S, \theta_K, \phi_K, D, \chi_0, \psi_0, \phi_0). 
\end{equation}
where $\chi_0,\psi_0,\phi_0$ are the initial values of $\chi,\psi,\phi$. 

The five parameters $(r_{\rm p}, M, m, a, D)$ can be obtained during the Monte Carlo sampling from the catalogs of simulated events introduced in Sec.\ref{paralist}. The other parameters to construct EMRB waveforms are: The sky position of the sources $(\theta_S, \phi_S)$ and the spin of the MBHs $(\theta_S, \phi_S)$, which we draw from an isotropic distribution on the sphere, and the inclination angular $\iota$, which we assume to be uniformly distributed in $(0, \pi)$. As EMRB has a parabolic-like orbit, we assume the CO orbit evolved with the mean anomaly $\phi$ range from $-\pi$ to $\pi$, and set $(\chi_0,\psi_0,\phi_0)$ in the middle of the orbit, with $\psi_0=0$, $\chi_0=0$ and $\phi_0=0$.  We point out that the CO can rotate around the MBH on a prograde and retrograde orbit. When $\iota<\pi/2$, CO will have a prograde orbit with $L_z>0$. When $\iota>\pi/2$, CO will have retrograde orbit with $L_z<0$. 

\begin{figure}[htbp]
	\centering
	\begin{minipage}{0.45\linewidth}
		\centering
		\includegraphics[width=\linewidth]{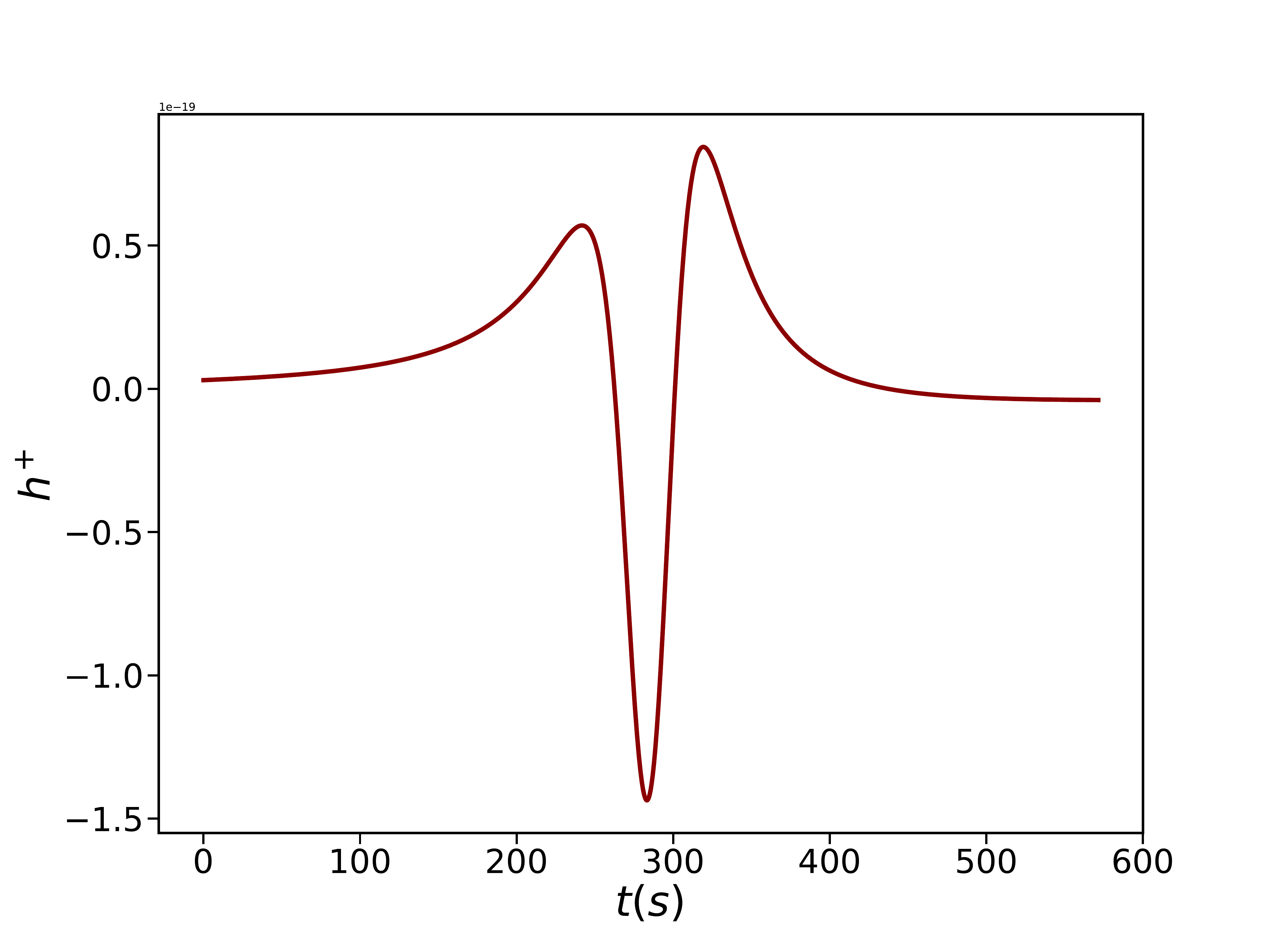}
		\label{paraM}
	\end{minipage}
	\begin{minipage}{0.45\linewidth}
		\centering
		\includegraphics[width=\linewidth]{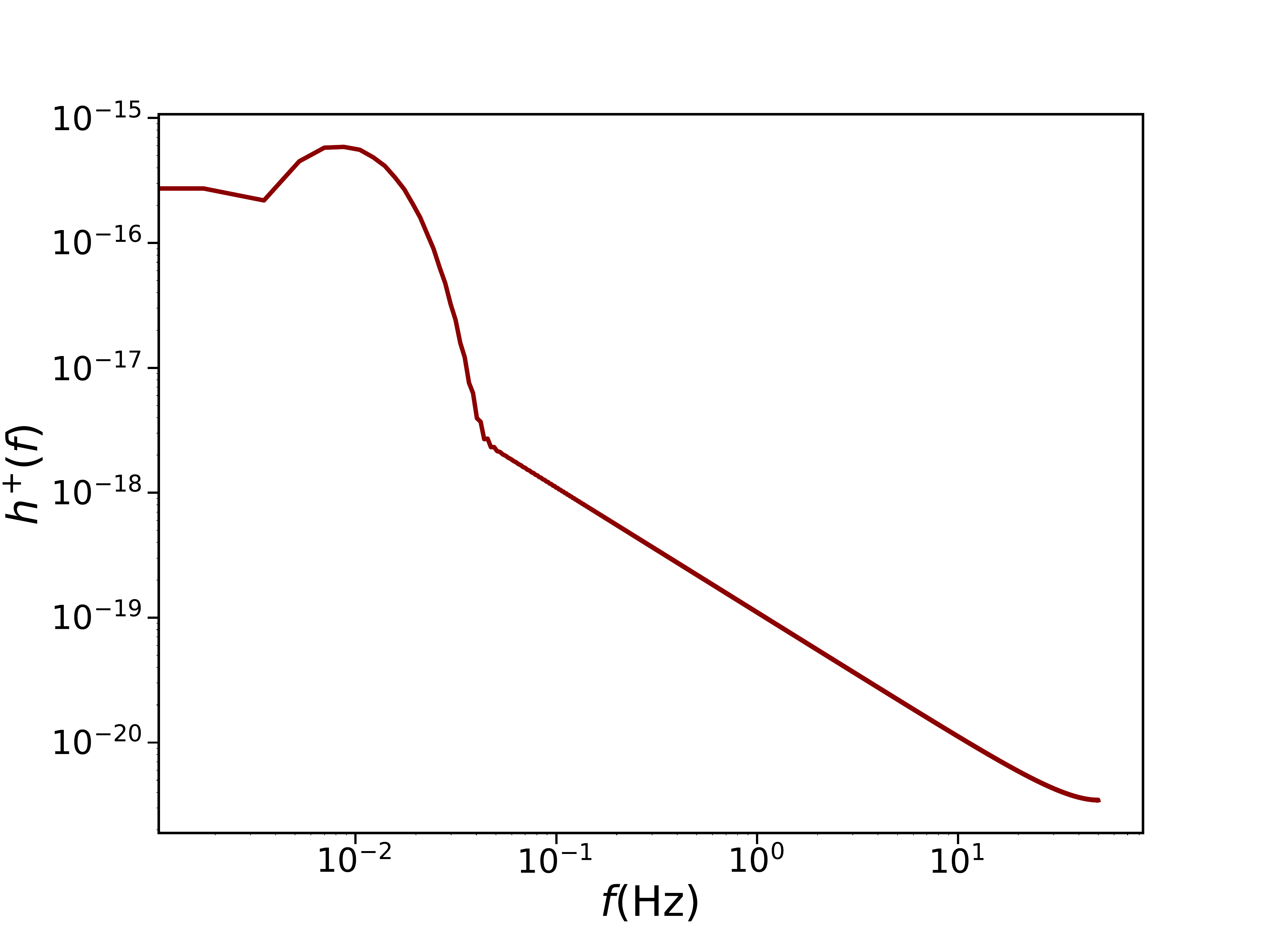}
		\label{paraa}
	\end{minipage}
		\begin{minipage}{0.45\linewidth}
		\centering
		\includegraphics[width=\linewidth]{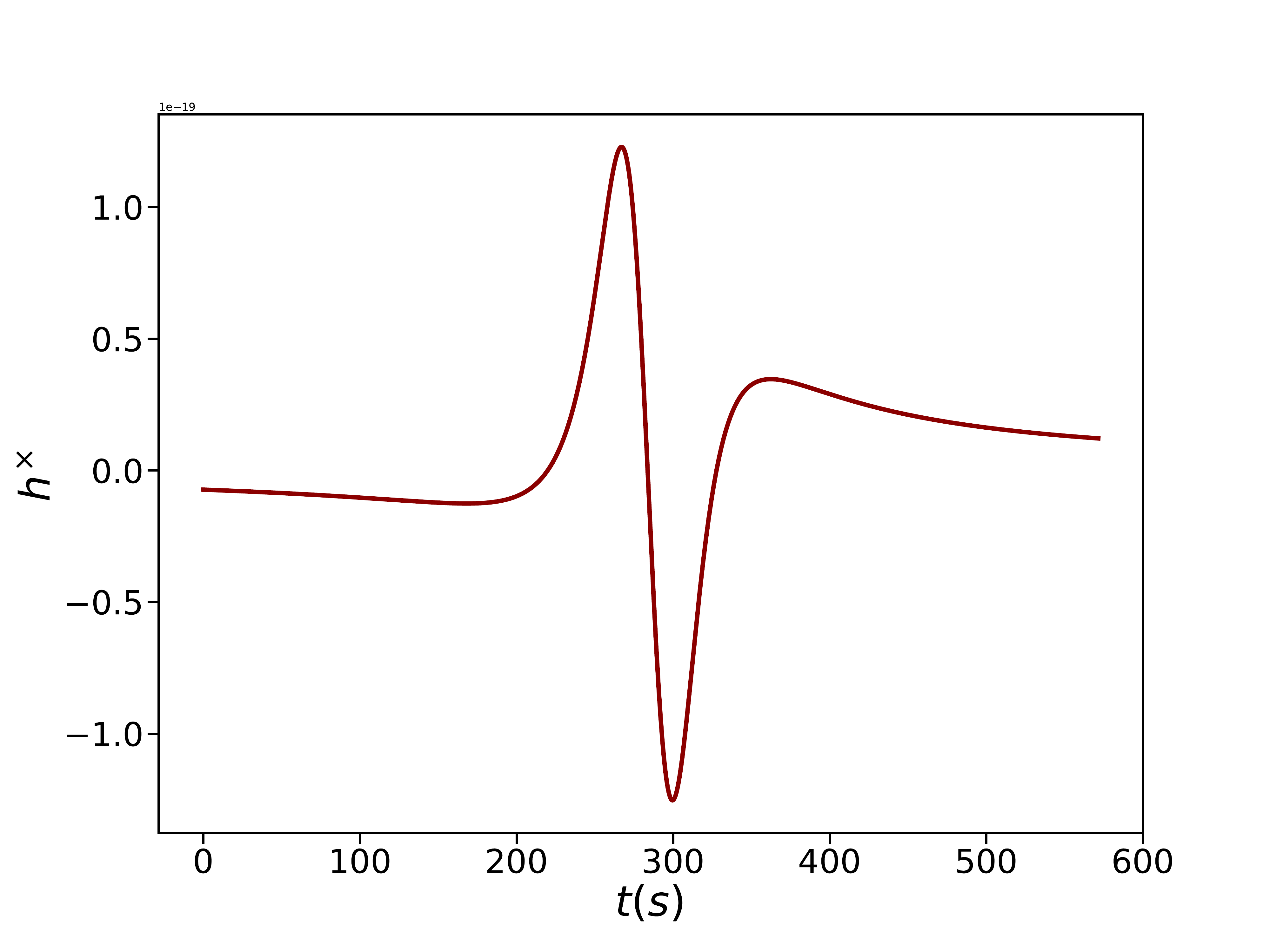}
		\label{paraM}
	\end{minipage}
	\begin{minipage}{0.45\linewidth}
		\centering
		\includegraphics[width=\linewidth]{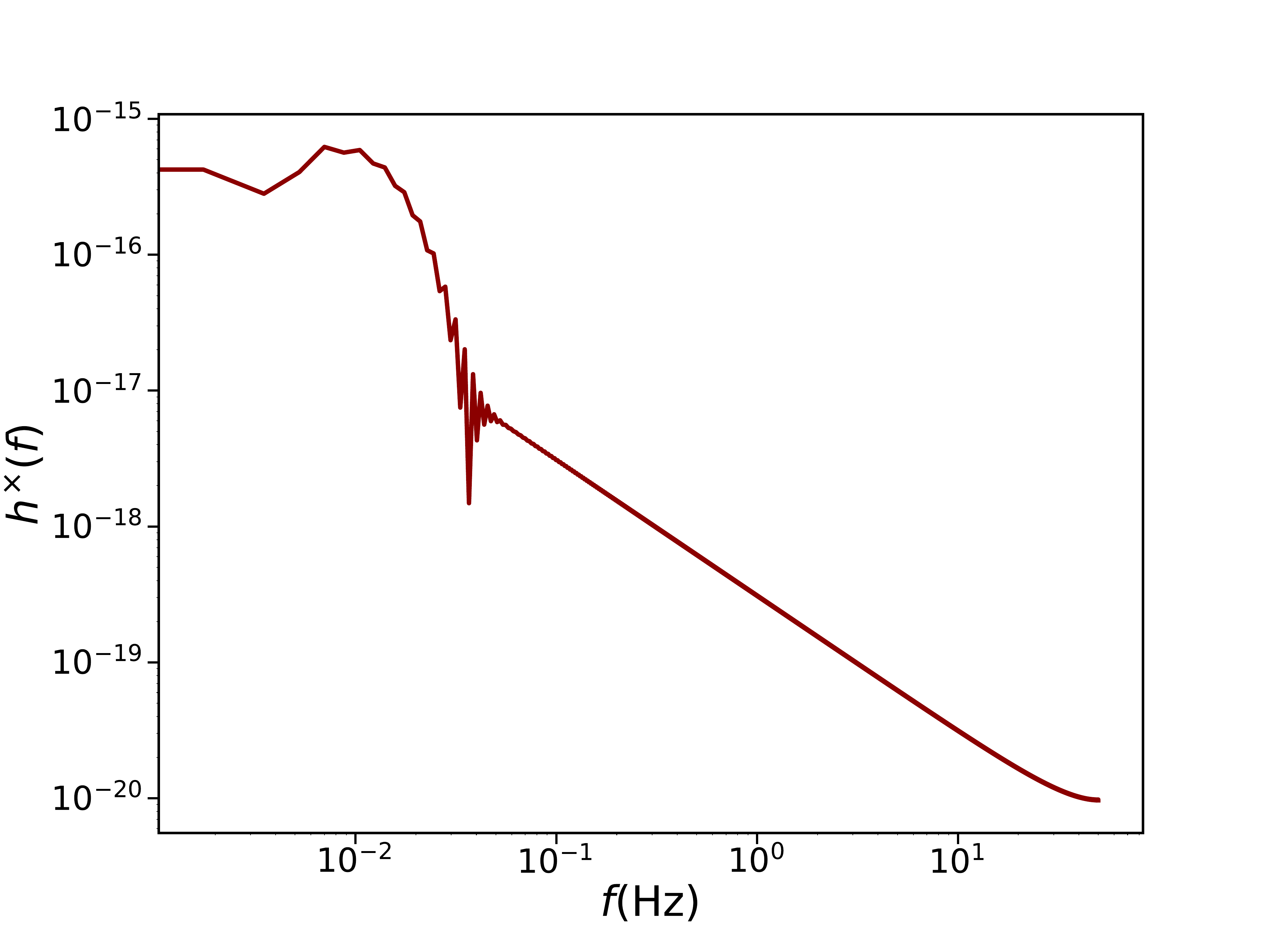}
		\label{paraa}
	\end{minipage}
	
\caption{The EMRB waveform in the time domain and frequency domain. }
\label{Waveform}
\end{figure}

Fig.\ref{Waveform} shows the waveform in  time domain and frequency domain for an EMRB source with $M=3\cdot10^5M_\odot, m=10M_\odot, r_p=6M, e=0.98, D=$1Mpc.

An EMRB traveling in direction $\hat{n}$ has a strain amplitude in the TianQin detector, which can be written as
\begin{equation}
h(t)=F^+(t)h^+(t)+F^\times(t)h^\times(t).
\end{equation}
Here, $F^+,F^\times$ are the functions\cite{Fan:2020zhy,Rubbo:2003ap,Cornish:2002rt}
\begin{equation}
\begin{split}
F^+(t)&=\cos(2\Psi)D^+(t)-\sin(2\Psi)D^\times(t),\\
F^\times(t)&=\sin(2\Psi)D^+(t)+\cos(2\Psi)D^\times(t),\\
\end{split}
\end{equation}
where $\Psi$ is the polarization angle, and $D_+(t)$ and $D_\times(t)$ describe the detector tensor. In the low frequency limit, where $f<f_*\approx0.28$Hz, the $D_+(t),D_\times(t)$ for TianQin have detailed expressions in \cite{Feng:2019wgq}.

\section{method}
\subsection{signal-to-noise ratio}
The signal-to-noise ratio (SNR) of the wave is defined using the noise-weighted product function\cite{Finn:1992wt}
\begin{equation}
(s_1|s_2)=2\int^\infty_0\frac{\tilde{s}_1(f)\tilde{s}^*_2(f)+\tilde{s}^*_1(f)\tilde{s}_2(f)}{S_n(f)}df,
\end{equation}
where $\tilde{s}_{1,2}$ are the signals, and $S_n(f)$ is the power spectral density (PSD) of the detector.  Then, the SNR can be described by
\begin{equation}
\rho=(h|h)^{1/2}=2\Big[\int^\infty_0\frac{\tilde{h}(f)\tilde{h}^*(f)}{S_n(f)}df\Big]^{1/2},
\end{equation}
Here, in our calculation, $h(f)$ is the EMRB signals, $S_n(f)$ is the PSD of TianQin detectors\cite{TianQin:2015yph}. 
TianQin has three arms, from which two independent Michelson signals $h_I$, $h_{II}$ can be constructed.  The total SNR of the signal can then be computed as
\begin{equation}
\rho=\sqrt{\rho_I^2+\rho_{II}^2}.
\end{equation}
The mission lifetime of TianQin is 5yrs, and its observation scheme is `` 3 months on +3 months off ".  As the typical duration of  EMRB signals is $\sim10^3$s, which is far smaller than 3 months, we assume each EMRB can be detected completely without any loss of information, while we expect to only detect half of the events enter in the TianQin band.

\subsection{Fisher information matrix}
The Fisher information matrix (FIM) is a common tool used to quantify the parameter measurement uncertainty\cite{Vallisneri:2007ev}\cite{Rodriguez:2013mla}. Here, we use FIM to derive a covariance matrix, whose diagonal value represent the estimation precision for an unbiased physical parameter.

 The FIM matrix is defined as,
\begin{equation}
\Gamma_{ij}=\Big(\frac{\partial{\tilde{h}(f)}}{\partial\lambda^i}\Big|\frac{\partial{\tilde{h}(f)}}{\partial\lambda^j}\Big)
\end{equation}
where $\lambda_i, i=1,2,...,$ are the parameters of the EMRB. The covariance matrix can be obtained as,
\begin{equation}
\Sigma_{ij}\equiv\langle\delta\lambda_i\delta\lambda_j\rangle=(\Gamma^{-1})_{ij}
\end{equation}
Then the uncertainty $\sigma_i$ for the $i$th parameter can be derived as,
\begin{equation}
\sigma_{i}=\Sigma^{1/2}_{ii}
\end{equation}

For the sky localization, it is convenient to use the solid angle $\Delta\Omega$ which corresponds to an error ellipse. Its can be expressed as a combination of the uncertainties on the ecliptic longitude angle $\phi_S$ and the ecliptic latitude angle $\theta_S$,
\begin{equation}
\Delta\Omega=2\pi|\sin\theta_S|\sqrt{\Sigma_{\theta_S}\Sigma_{\phi_S}-\Sigma^2_{\theta_S\phi_S}}
\end{equation}

\section{Results}\label{sec:detrate}

\subsection{detection number}

\begin{figure}
\centering
\includegraphics[width=0.85\linewidth,clip=true,angle=0,scale=0.75]{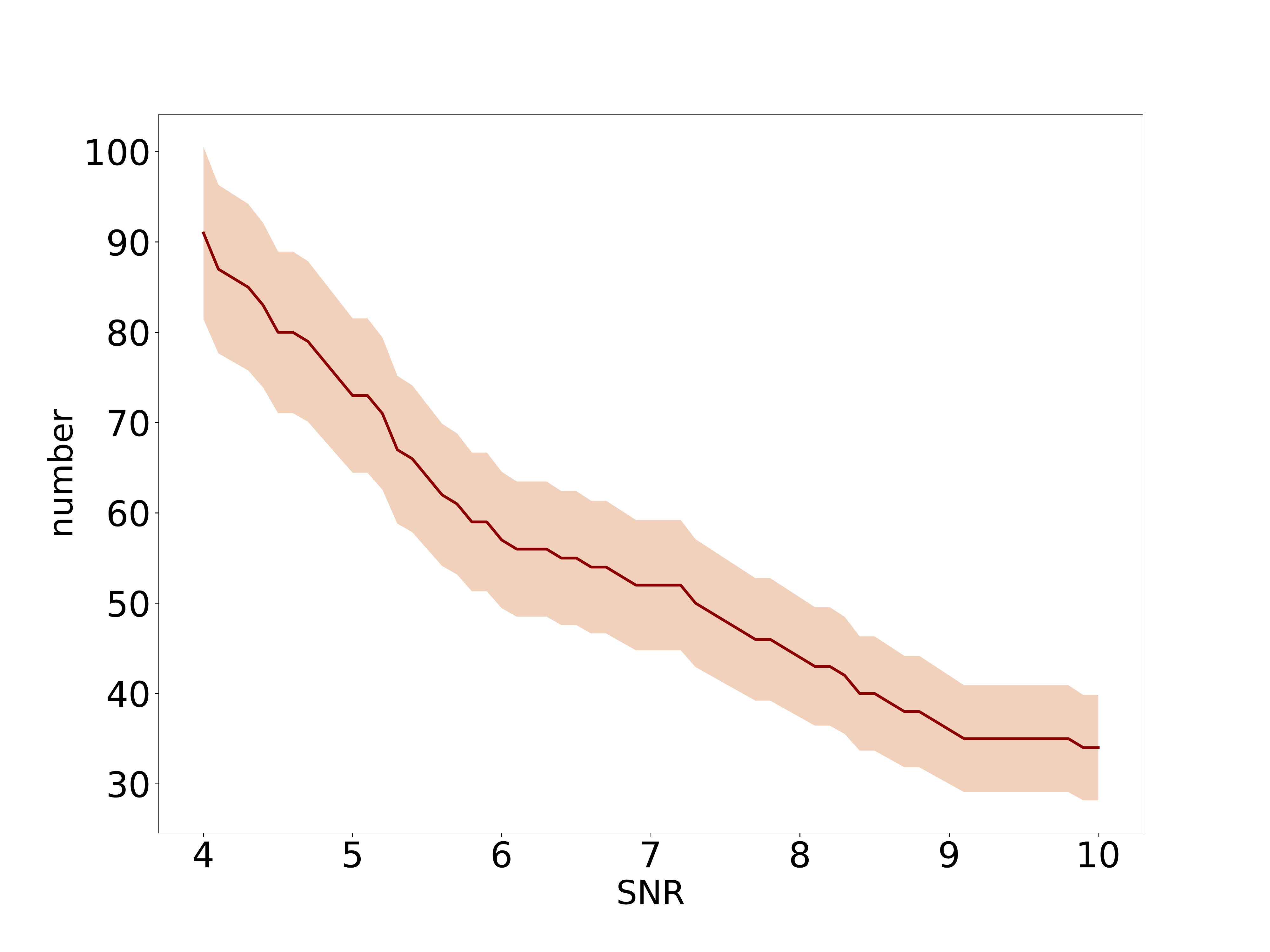}
\caption{The darkred line corresponds to the EMRB events number distribution with SNR above 4 and the shadow corresponds to the uncertainty assuming a Poisson distribution. }
\label{fig:snrdistribut}
\end{figure}

EMRBs can provide useful information to the study of astronomy and astrophysics. Those EMRB events with SNR larger than 10 are usually considered to be well identified during the data analysis\cite{Berry:2013poa}. Here, we calculate the SNR for each EMRB event, finding TianQin can detect about 35 EMRB events during its mission lifetime.  As there will be multiple space-borne GW detectors in the future, the GW signals can be mutual verified in different detectors and the EMRB detection threshold may be greatly decreased. Here, we present the number of EMRB signals with a SNR larger than 4 in Fig.\ref{fig:snrdistribut}. In this figure, the darkred line corresponds to the EMRB events number distribution and the shadow corresponds to the uncertainty assuming a Poission distribution.

\begin{figure}
\centering
\includegraphics[width=0.85\linewidth,clip=true,angle=0,scale=0.75]{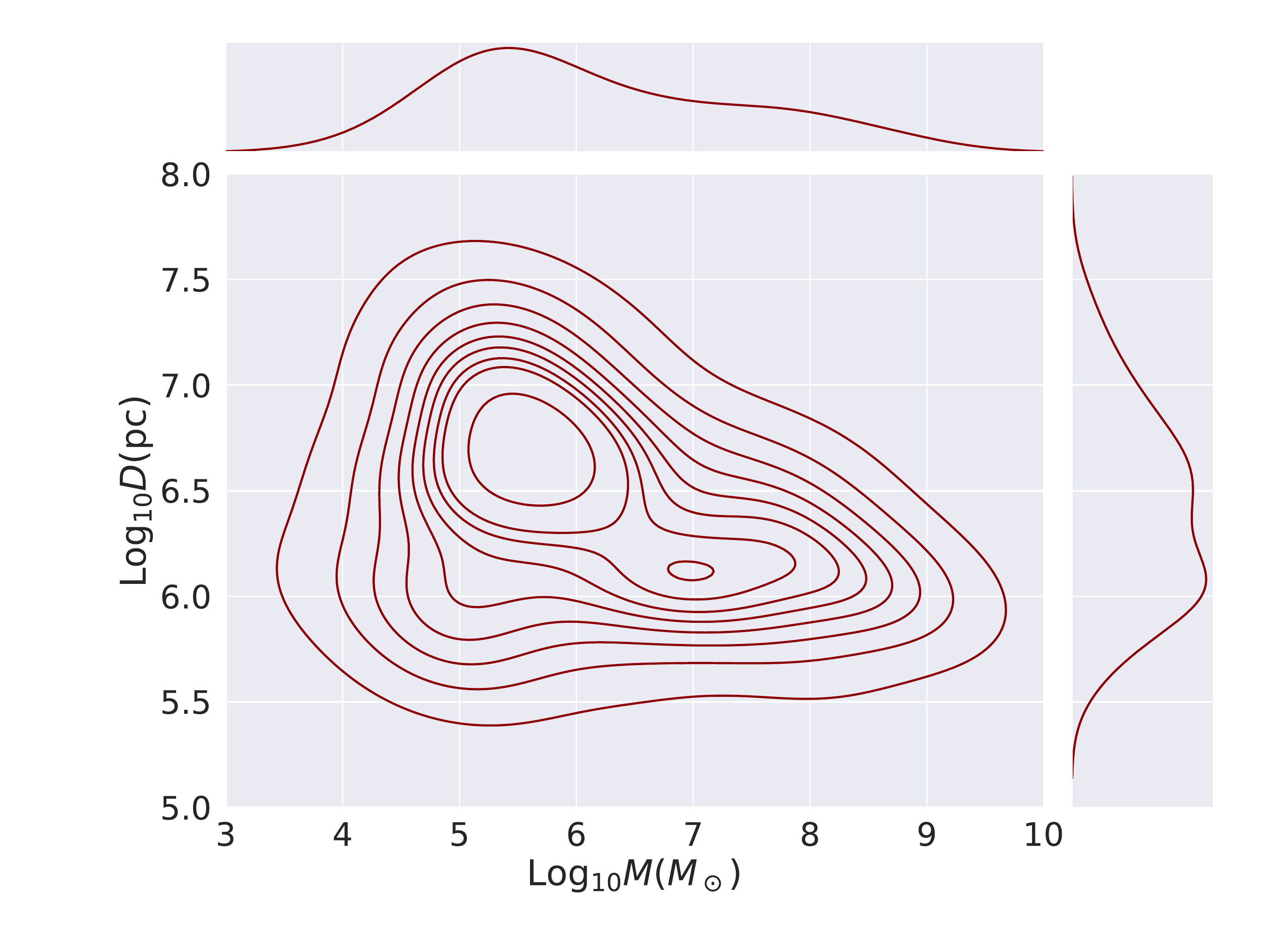}
\caption{The strong EMRB events distribution with $M$ and $D_L$. }
\label{fig:eventdistribut}
\end{figure}

For those strong signals with SNR larger than 10, we display their distribution with different massive black hole mass and luminosity distance in Fig.\ref{fig:eventdistribut}. Our result shows that MBH with a mass between $10^5\sim10^6M_\odot$ is more likely to be detected, which is consistent with the result for EMRIs detection by TianQin. As we explained in \cite{Fan:2020zhy}, this feature is mostly related to the frequency dependence of the TianQin sensitivity curve and the relation between the MBH mass and the frequency of a GW signal. From this figure, we can also find that the horizon distance of EMRB for TianQin can be set to 100Mpc, as no EMRB events with a distance above this value were detected during our calculation. The other relevant parameters are the CO mass, the pericenter distance and the MBH spin.  The effect of CO mass and pericenter distance on EMRB is obvious since increasing the first and decreasing the second will enhance the detecability of EMRB. We also explore the event distribution for different MBH spin, finding that  the  effect of the spin related to the detection number is not consistent with result for EMRIs. For EMRIs,  MBHs with higher spin are easier to detect because they generate EMRIs with smaller last stable orbit (LSO), thus resulting in a larger amplitude of the wave. However, for EMRB, the MBH spin has no significant effect on detection number as the pericenter orbit is already fixed during the Monte Carlo sampling.

Beside being interesting target sources, EMRB signals appearing in the detector can also affect the detection and analysis of other sources.  In particular, EMRB can be confused with glitches and identified as correlated noise existing in the detected data\cite{LIGOScientific:2016gtq}.  In fact, there exists one possible method to separate the burst signals, which is the \textit{null channel}\cite{Goncharov:2022dgl}. In that method, data analysis is performed on the TDI channels, which include A, E, and T channels. Among them, A and E channels are conventional channels, while the T channel is insensitive to GW signals and can be used to identify glitches and burst signals. However, how strong a burst signal need to be so that it can be well separated using this method requires more detailed studies.

\subsection{Parameter estimation}

As we have said, the GW detection of EMRB sources can provide an opportunity to improve our fundamental understanding of the evolution of the MBHs and help us gain a glimpse into the past of their host galaxies. Here, we use the FIM to investigate the parameter estimation accuracy of these sources. 

\begin{figure}[htbp]
	\centering
	\begin{minipage}{0.495\linewidth}
		\centering
		\includegraphics[width=\linewidth]{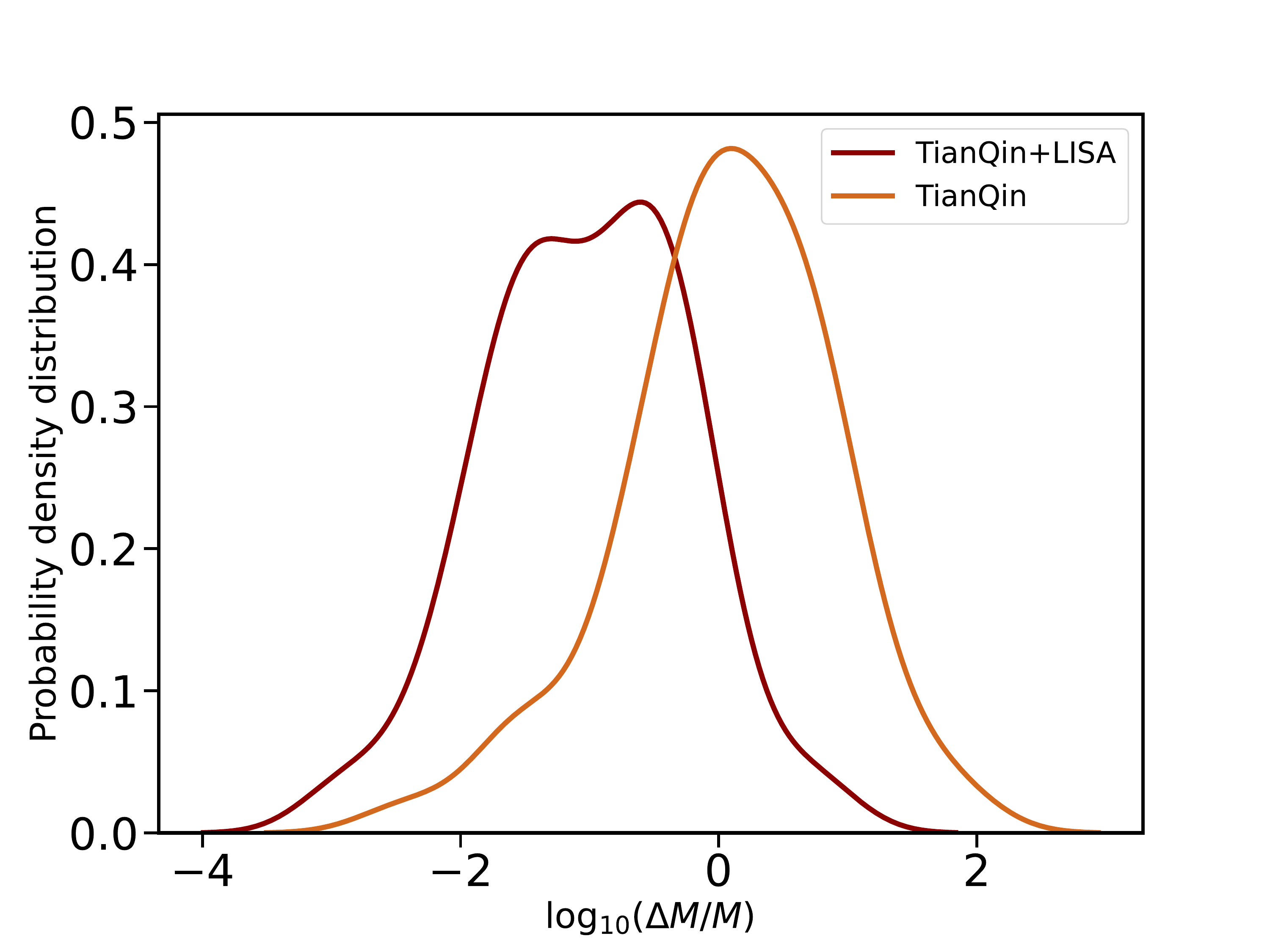}
		\label{paraM}
	\end{minipage}
	\begin{minipage}{0.495\linewidth}
		\centering
		\includegraphics[width=\linewidth]{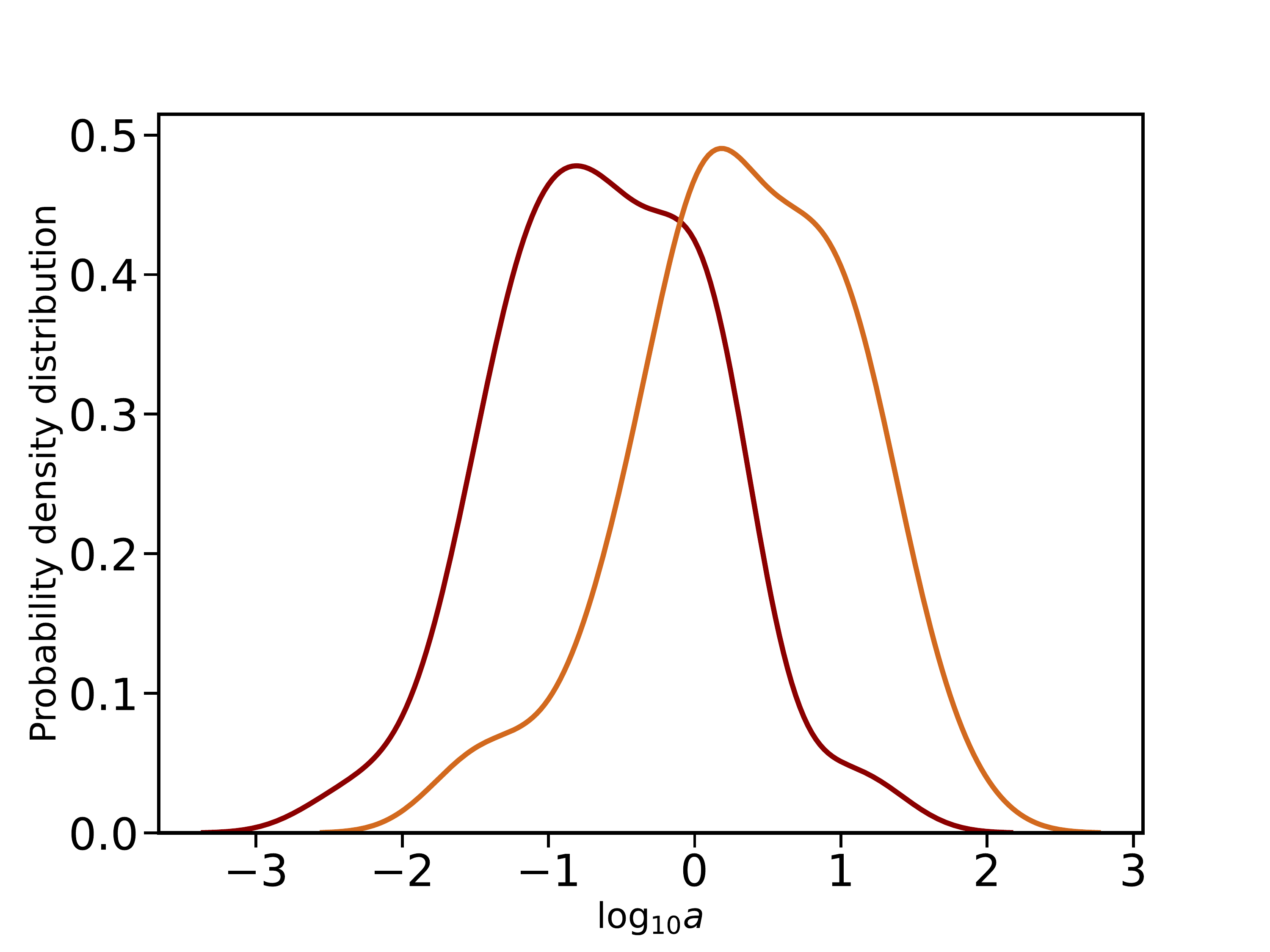}
		\label{paraa}
	\end{minipage}

	\begin{minipage}{0.495\linewidth}
		\centering
		\includegraphics[width=\linewidth]{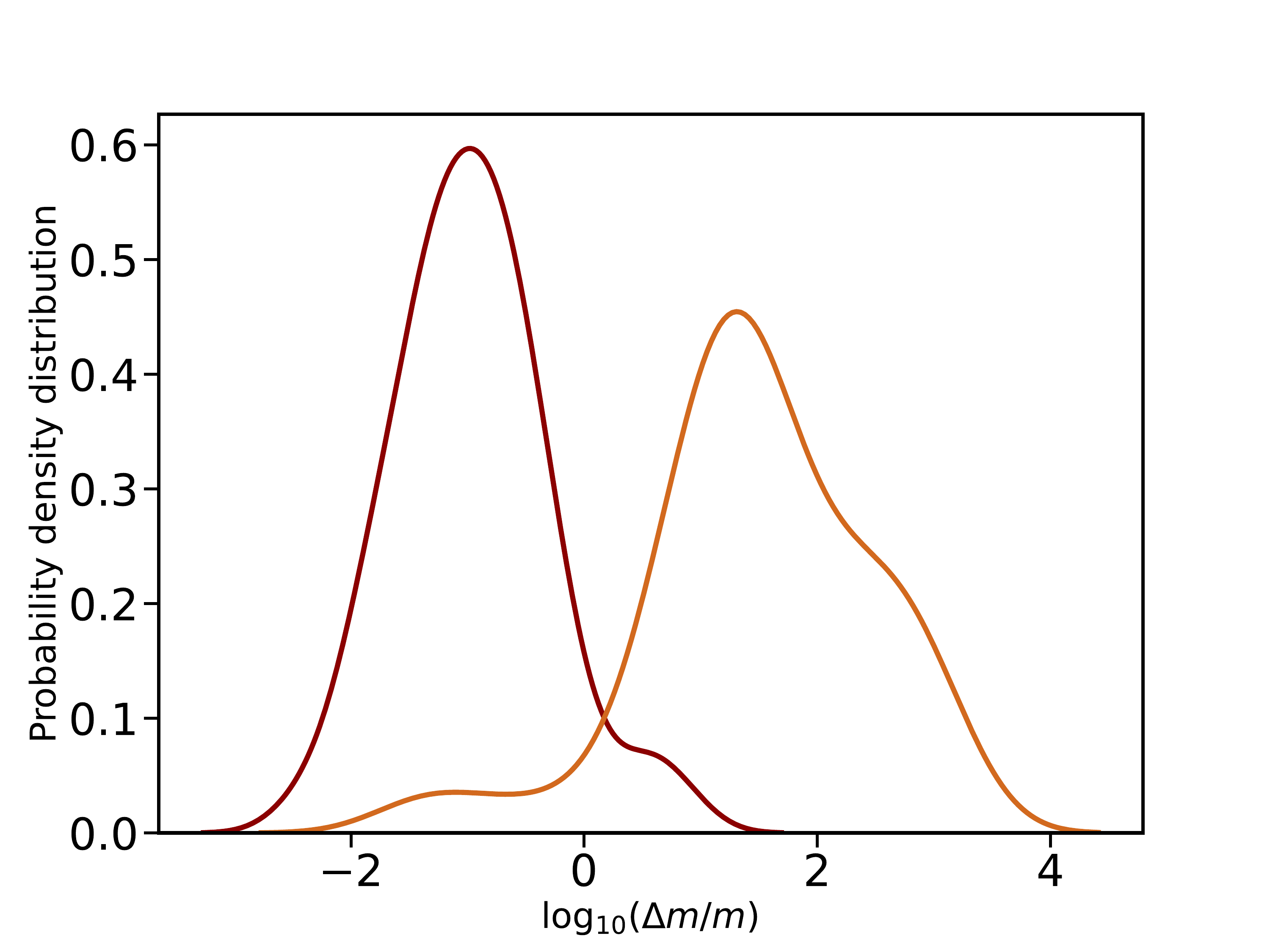}
		\label{param}
	\end{minipage}
	\begin{minipage}{0.495\linewidth}
		\centering
		\includegraphics[width=\linewidth]{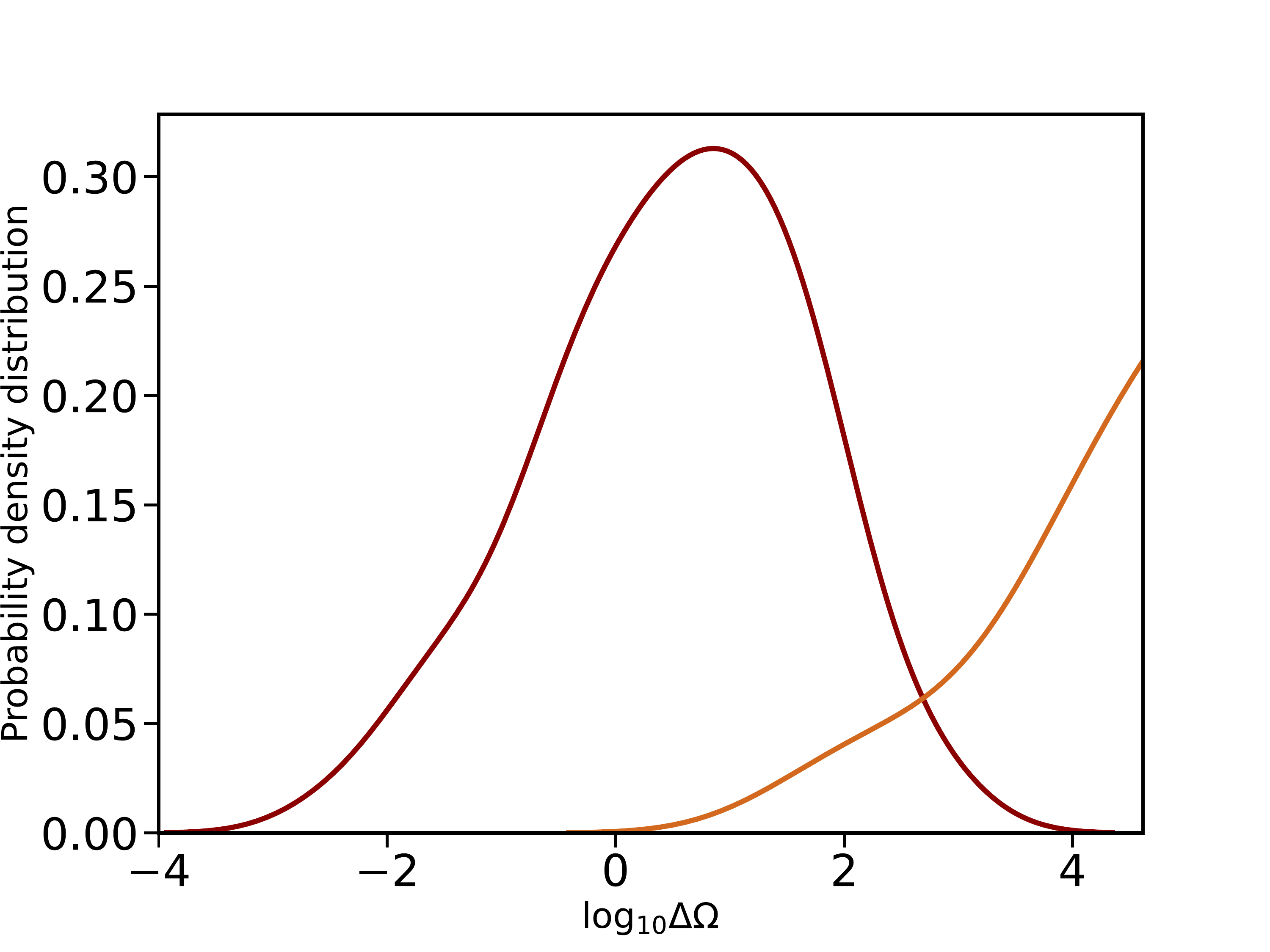}
		\label{paraOme}
	\end{minipage}
\caption{The parameter estimation precision for various parameters by TianQin (yellow) and by TianQin+LISA (darkred). }
\label{ParaEstimate}
\end{figure}

EMRBs are very short and hence their physical quantities $(E, L_z, Q)$ can be regarded as constant. This, however, will also cause the parameter degeneracy between the CO mass $m$ and the luminosity distance $D_L$.  A common method to handle the degeneracy is to fix $m$ or $D_L$ and then to remove the fixed parameter while calculating the FIM for the remaining parameters.  In our work, we choose to fix the luminosity distance,  because EMRB sources are ususally close, if their spatial position can be located precisely, their luminosity distance may be determined through electromagnetic observations. 
EMRBs have 13 parameters, of which we are more interested in $M, m, a, \theta_S, \phi_S$, where $\theta_S, \phi_S$ determine the sky localization $\Delta\Omega$. For those sources detected with a SNR larger than 10, we present the parameter estimation accuracy  in Fig.\ref{ParaEstimate}.

In Fig.\ref{ParaEstimate}, we see that, in the vest case, TianQin can determine the MBH mass, the MBH spin and the CO mass with an accuracy of $10^{-2}$, and determine the sky localion with an accuracy of  10 square degree. However, for the general cases, those EMRB source parameters can't be determined very well. As there will be multiple space-borne detectors in the future, their joint detection will provide more information on the EMRB and improve their parameter estimation accuracy. We consider the case of multidetectors TianQin+LISA and analyse the impact on parameter estimation. Our result shows that, using TianQin+LISA, the estimation accuracy of EMRB parameters can be improved greatly. For the most  precise case, the MBH mass, the MBH spin and the CO mass can be determined with an accuracy of $10^{-3}$ and the sky location can be determined with an accuracy of $10^{-2.5}$ square degree. For the general case, the MBH mass, the MBH spin, and the CO mass can be determined with an accuracy of $10^{-1}$, while the sky localization can be determined within 10 square degrees.



\subsection{Gravitational Wave Background}
Besides the detectable EMRB signals, there is a large fraction of EMRB events will not be individually resolvable due to their relatively weak signal. Their incoherent superposition constitute a source of gravitational wave background (GWB). In this paper.  We estimate the shape and the overall magnitude of the GWB generated by EMRBs and show its detectability by TianQin.  
 
As mentioned above, we set the detection threshold for EMRB to be 10.  After removing these strong signals, we calculate the GWB generated by the remaining events. Following Sec.\ref{sec:wave}, we can calculate the GW energy released by a single EMRB and estimate its spectrum. By combining this spectral shape with the list of faint EMRB sources, the spectrum of GWB can be calculated. 
By summing over the list of each spectrum of EMRB events and then getting its average value, the GWB is obtained as
\begin{equation}
\langle\tilde{h}(f),\tilde{h}^*(f^\prime)\rangle=\frac{\sum_i{\tilde{h}_i(f){\tilde{h}^*_i(f)}}}{T_{\rm det}},
\end{equation}
where $\tilde{h}(f)$ is the Fourier transform of $h(t)$, $i$ is the $i$-th EMRB event and $T_{\rm det}$ is the detector observation time.

The resulting GWB of EMRBs is shown in Fig.\ref{fig:fmrp}. In this figure, the yellow line represents the power-law sensitivity curve of TianQin,  and the red line represents the GWB generated by the EMRB events. From this figure, we can see that the GWB generated by EMRBs is about $10^6$ times weaker than TianQin's sensitivity and thus can be ignored during TianQin detection. 
 \begin{figure}
\centering
\includegraphics[width=\columnwidth,clip=true,angle=0,scale=0.75]{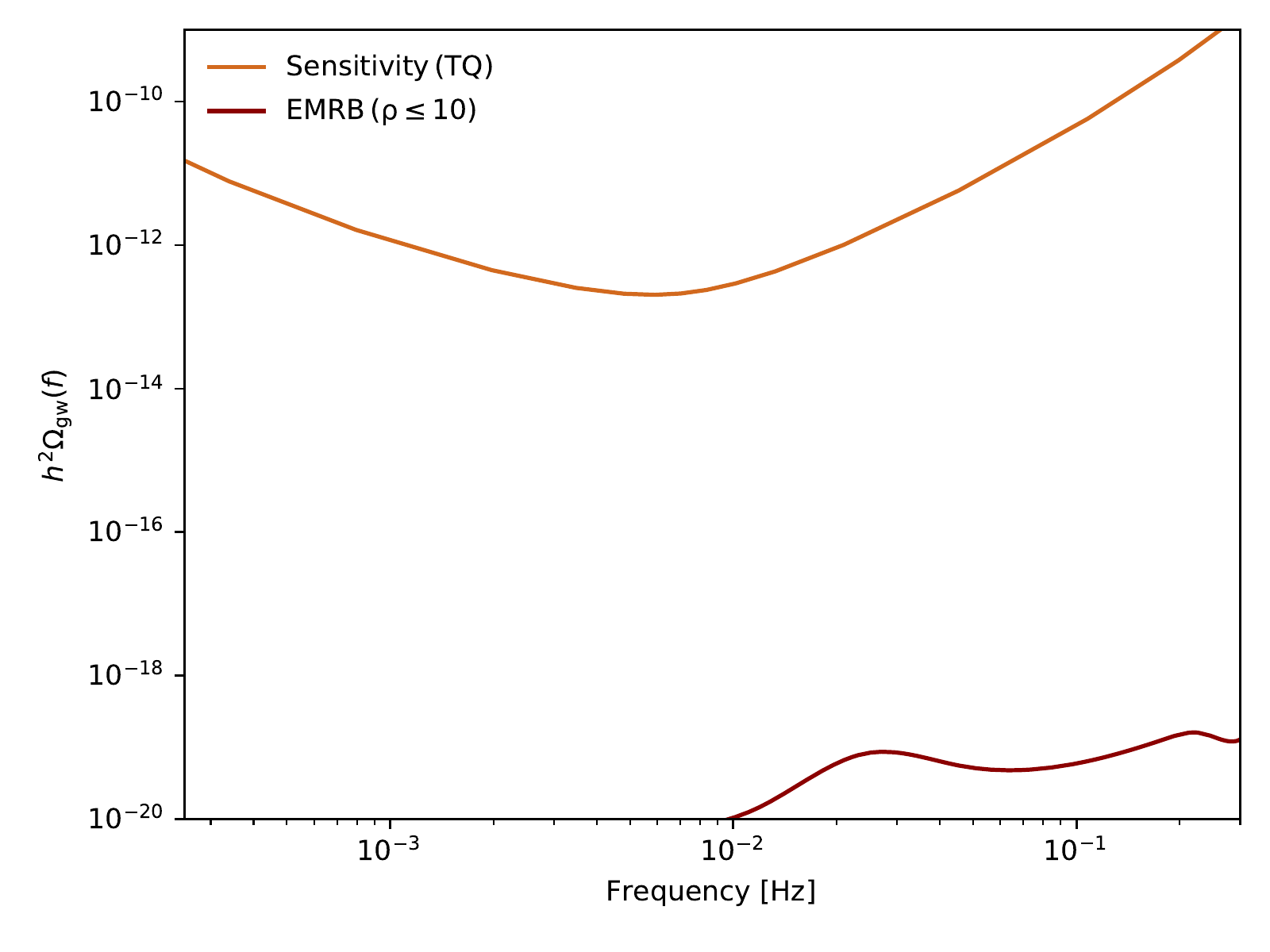}
\caption{The black line is the power-law sensitivity curve of TianQin, in which we assume that the operation time $T_{\rm op}=1\,\,{\rm yr} $and the SNR thredshold $\rho_{\rm thr}=1$. The red line is the $\Omega_{\rm gw}$ of the EMRB, which are the sources with a SNR of less than 10 after one year of TianQin's operation. }
\label{fig:fmrp}
\end{figure}



\section{Conclusion and Discussion}\label{sec:sum}

In this work, we perform a preliminary study of the detectability of TianQin on EMRBs. We use direct N-body simulations to study the rate of EMRBs depending on the mass of MBH and extend this result to get the EMRB distribution.  For each EMRB, we use a method based on NK to obtain its waveform.

Our results show that, choosing the detection threshold as 10, we expect that tens of EMRBs can be identified during the TianQin mission lifetime. We also find that those EMRBs with a MBH mass between $10^5\sim10^6M_\odot$ are more likely to be detected and that the furthest detectable distance for EMRBs is no more than 100Mpc.

We calculate the expected precision of the parameter estimation using the FIM method. By using the TianQin+LISA to improve the precision, our results show that, for most of the sources, the MBH mass, the MBH spin and the CO can be determined with an accuracy of $10^{-1}$, and the luminosity distance can be determined with an accuracy of 10 square degree.

We further study the GWB generated by unresolved EMRBs, finding that it is smaller than the PSD of TianQin by about 6 orders, thus it can be ignored.

\section{Acknowledgements} 
We are grateful to Alejandro Torres Orjuela for his helpful discussion. This work has been supported by Guangdong Major Project of Basic and Applied Basic Research (Grant No.2019B030302001).
~\\

\bibliographystyle{iopart-num}
\bibliography{reference}

\end{document}